\documentclass[a4paper,10pt]{article}
\usepackage{amssymb}
\usepackage{amsmath}
\usepackage{latexsym}
\usepackage{array}
\usepackage{graphicx}

\def\DIS{\displaystyle}
\def\hbreak{\vspace*{3mm}\hfill\break\noindent}
\usepackage{theorem}
\theoremstyle{break}
\newtheorem{Theorem}{Theorem}
\newtheorem{Proposition}{Proposition}
\newtheorem{Remark}{Remark}

\def\maru#1{{#1}^\cap}

\def\Z{{\mathbb Z}}

\begin{document}
\title{On the initial value problem of a periodic box-ball system}
\author{Jun Mada${}^1$, Makoto Idzumi${}^2$ and Tetsuji Tokihiro${}^1$ \\ \\
${}^1$ Graduate school of Mathematical Sciences, \\
      University of Tokyo, 3-8-1 Komaba, Tokyo 153-8914, Japan \\
${}^2$ Department of Mathematics, Faculty of Education, \\
      Shimane University, Matsue 690-8504, Japan}

\date{}

\maketitle

\begin{abstract}
We show that the initial value problem of a periodic box-ball system
can be solved in an elementary way using simple combinatorial methods.
\end{abstract}


A periodic box-ball system (PBBS) is a dynamical system of balls in 
an array of boxes with a periodic boundary condition \cite{TS,YT}.
The PBBS is obtained from the discrete KdV equation and the discrete 
Toda equation, both of which are known as typical integrable nonlinear discrete equations, 
through a limiting procedure called ultradiscretization \cite{TTMS,MSTTT}. 
Since the ultradiscretization preserves the main properties of the original discrete equations, 
and the solvability of the initial value problem being an important property of integrable equations, 
we expect that the initial value problem of the PBBS can also be solved.
In fact, the initial value problem for the PBBS was first solved by inverse ultradiscretization combined 
with the method of inverse scattering transform of the discrete Toda equation \cite{KT} 
and recently by the Bethe ansatz for an integrable lattice model with quantum group symmetry 
at the deformation parameter $q=0$ and $q=1$ \cite{KTT}.
These two methods, however, require fairly specialized mathematical knowledge 
on algebraic curves or representation theory of quantum algebras. 

An important property which characterizes a state of the PBBS is the fundamental cycle of the state, 
${i.e.}$, the length of the trajectory to which it belongs.
Its explicit formula as well as statistical distribution was obtained 
and its relation to the celebrated Riemann hypothesis was clarified \cite{YYT,MT,TM}.
To prove the formula for fundamental cycle, one of the key steps is to compare a state 
with its `reduced states' constructed by the `10-elimination'.
In this article, we show that the initial value problem of the PBBS is solved by simple combinatorial arguments 
-- essentially given in Ref.~\cite{YYT} -- with some remarkable features of the reduced states. 

First we quickly review the definition of the PBBS and its conserved quantities.
Consider a one-dimensional array of boxes each with a capacity of one ball. 
A periodic boundary condition is imposed by assuming that the last box is adjacent to the first one.
Let the number of boxes be $N$ and that of balls be $M$. We assume $M<N/2$. 
An arrangement of $M$ balls in $N$ boxes is called a state of the PBBS. 
Denoting a vacant box by $0$ and a filled box by $1$, 
a state of the PBBS is represented by a $0,1$ sequence of length $N$. 
The time evolution rule from time step $t$ to $t+1$ can be described as follows: 
\begin{itemize}
\item For a given state, connect all $10$ pairs in the sequence with arc lines. 
      We call them `$\maru{1}$arc lines'.
\item Neglecting the $10$ pairs which are connected in the first step, 
      connect all the remaining $10$s with arc lines. 
      We call them `$\maru{2}$arc lines'.
\item Repeat the above procedure until all $1$s are connected to $0$s with arc lines.
\item Exchange all the $1$s and $0$s which are connected with arc lines. 
      Then we obtain a new sequence which we call the state evolved by one time step. 
\end{itemize}
If we denote by $p_j(t)$ the number of $\maru{j}$arc lines, 
we obtain a nonincreasing sequence of positive integers, $p_j(t)\ (j=1,2,3,\ldots,m)$. 
Then, this sequence is conserved in time, that is, 
\begin{equation*}
 p_j(t)=p_j(t+1)\equiv p_j\qquad (j=1,2,3,\ldots,m).
\end{equation*}
As the sequence $(p_1,p_2,\ldots ,p_m)$ is nonincreasing, 
we can associate a Young diagram to it by regarding $p_j$ as the number of 
squares in the $j$th column of the diagram. 
The lengths of the rows are also weakly decreasing positive integers. 
Let the distinct row lengths be $L_1>L_2> \cdots >L_s$ 
and let $n_j$ be the number of times that length $L_j$ appears. 
The set $\DIS \{L_j, n_j\}_{j=1}^s$ is another expression 
for the conserved quantities of the PBBS.

\begin{figure}[hbpt]
 \begin{center}
     \includegraphics[scale=0.5]{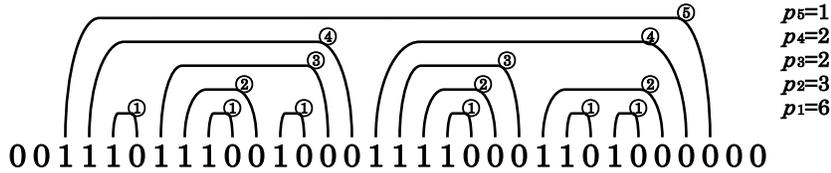}
 \end{center}
  \caption{Arc lines and conserved quantities for the state $(\sharp)$.}
  \label{fig:Rule}
\end{figure}

\begin{figure}[hbpt]
 \begin{center}
  \includegraphics[scale=0.5]{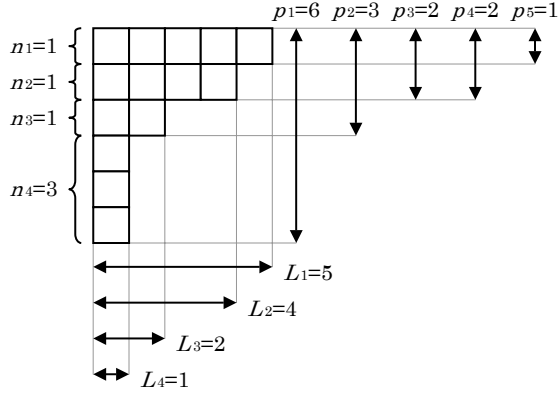}
 \end{center}
  \caption{Young diagram corresponding to the conserved quantities of $(\sharp)$.}
  \label{YoungD}
\end{figure}

For example, for a state of the PBBS with $N=32,\ M=14$ 
\begin{equation*}
 (\sharp) \qquad 00111011100100011110001101000000, 
\end{equation*}
the arc lines are drawn as in Fig.~\ref{fig:Rule} and its conserved quantities are
expressed by the Young diagram given in Fig.~\ref{YoungD}. 

To solve the initial value problem of the PBBS, we utilize the fact that a state of the PBBS 
is determined by its conserved quantities and the positions of $10$ pairs 
to which the `$10$-elimination' is applied.
In the present context, a $10$-elimination is to convert a state to a state 
with smaller number of entries by eliminating all $10$ pairs in the sequence connected with $\maru{1}$arc lines.
Let us consider a state ${\cal S}$ with conserved quantities 
$(p_1,p_2,\ldots,p_m)$ or equivalently $\DIS \{L_j, n_j\}_{j=1}^s$. Note that $s \le m$ and $L_1=m$.
We define its $k$-reduced state ($k=1,2,\ldots, m$) as the one obtained 
from the state by eliminating all the $10$ pairs connected with $\maru{j}$arc lines for all $j:\ 1\le j \le k$. 
The length of the $k$-reduced state is $N-2\sum_{j=1}^k p_j$. 
We sometimes call the original state ${\cal S}$ as $0$-reduced state. 
Apparently $10$-elimination is not a reversible operation. 
If, however, we remember the places where $10$ pairs have been eliminated, 
we can recover the original state, at least up to shift, by inserting $10$ pairs there.
Since there has necessarily been a $10$ pair between consecutive $1$ and $0$ in the reduced state,
we have only to remember the places for the other $10$ pairs.
(Such places are called the positions of $0$-solitons in Ref.~\cite{YYT}, 
and we use the same terminology here.)
To specify the positions for insertion of $10$ pairs, 
we number a place between $j$ and $j+1$th consecutive entries in a state by integer $j$.
Note that, due to the periodic boundary condition, we set $0\equiv N-2\sum_{j=1}^k p_j$ 
for the $k$-reduced state.
Hereafter we explicitly write the $j$th place by `$\DIS \ \stackrel{j}{|}\ $' in the $0,1$ sequence.
For example, the state $(\sharp)$ is expressed as
\begin{equation*}
 \stackrel{0}{|}\hskip-0.8mm {0} \hskip-0.8mm 
 \stackrel{1}{|}\hskip-0.8mm {0} \hskip-0.8mm 
 \stackrel{2}{|}\hskip-0.8mm {1} \hskip-0.8mm 
 \stackrel{3}{|}\hskip-0.8mm {1} \hskip-0.8mm 
 \stackrel{4}{|}\hskip-0.8mm {1} \hskip-0.8mm 
 \stackrel{5}{|}\hskip-0.8mm {0} \hskip-0.8mm 
 \stackrel{6}{|}\hskip-0.8mm {1} \hskip-0.8mm 
 \stackrel{7}{|}\hskip-0.8mm {1} \hskip-0.8mm 
 \stackrel{8}{|}\hskip-0.8mm {1} \hskip-0.8mm 
 \stackrel{9}{|}\hskip-0.8mm {0} \hskip-1.4mm 
 \stackrel{10}{|}\hskip-1.4mm {0} \hskip-1.4mm 
 \stackrel{11}{|}\hskip-1.4mm {1} \hskip-1.4mm 
 \stackrel{12}{|}\hskip-1.4mm {0} \hskip-1.4mm 
 \stackrel{13}{|}\hskip-1.4mm {0} \hskip-1.4mm 
 \stackrel{14}{|}\hskip-1.4mm {0} \hskip-1.4mm 
 \stackrel{15}{|}\hskip-1.4mm {1} \hskip-1.4mm 
 \stackrel{16}{|}\hskip-1.4mm {1} \hskip-1.4mm 
 \stackrel{17}{|}\hskip-1.4mm {1} \hskip-1.4mm 
 \stackrel{18}{|}\hskip-1.4mm {1} \hskip-1.4mm 
 \stackrel{19}{|}\hskip-1.4mm {0} \hskip-1.4mm 
 \stackrel{20}{|}\hskip-1.4mm {0} \hskip-1.4mm 
 \stackrel{21}{|}\hskip-1.4mm {0} \hskip-1.4mm 
 \stackrel{22}{|}\hskip-1.4mm {1} \hskip-1.4mm 
 \stackrel{23}{|}\hskip-1.4mm {1} \hskip-1.4mm 
 \stackrel{24}{|}\hskip-1.4mm {0} \hskip-1.4mm 
 \stackrel{25}{|}\hskip-1.4mm {1} \hskip-1.4mm 
 \stackrel{26}{|}\hskip-1.4mm {0} \hskip-1.4mm 
 \stackrel{27}{|}\hskip-1.4mm {0} \hskip-1.4mm 
 \stackrel{28}{|}\hskip-1.4mm {0} \hskip-1.4mm 
 \stackrel{29}{|}\hskip-1.4mm {0} \hskip-1.4mm 
 \stackrel{30}{|}\hskip-1.4mm {0} \hskip-1.4mm 
 \stackrel{31}{|}\hskip-1.4mm {0} \hskip-1.4mm 
 \stackrel{32}{|}.
\end{equation*}

If we denote by $\hat{E}$ a $10$-elimination, for the above state $(\sharp)$, 
the $k$-reduced state is expressed as $\hat{E}^k(\sharp)$ and we have 
\begin{eqnarray*}
 \hat{E}(\sharp) &= & \stackrel{0}{|}\hskip-0.8mm {0} \hskip-0.8mm 
 \stackrel{1}{|}\hskip-0.8mm {0} \hskip-0.8mm 
 \stackrel{2}{|}\hskip-0.8mm {1} \hskip-0.8mm 
 \stackrel{3}{|}\hskip-0.8mm {1} \hskip-0.8mm 
 \stackrel{4}{|}\hskip-0.8mm {1} \hskip-0.8mm 
 \stackrel{5}{|}\hskip-0.8mm {1} \hskip-0.8mm 
 \stackrel{6}{|}\hskip-0.8mm {0} \hskip-0.8mm 
 \stackrel{7}{|}\hskip-0.8mm {0} \hskip-0.8mm 
 \stackrel{8}{|}\hskip-0.8mm {0} \hskip-0.8mm 
 \stackrel{9}{|}\hskip-0.8mm {1} \hskip-1.4mm 
 \stackrel{10}{|}\hskip-1.4mm {1} \hskip-1.4mm 
 \stackrel{11}{|}\hskip-1.4mm {1} \hskip-1.4mm 
 \stackrel{12}{|}\hskip-1.4mm {0} \hskip-1.4mm 
 \stackrel{13}{|}\hskip-1.4mm {0} \hskip-1.4mm 
 \stackrel{14}{|}\hskip-1.4mm {1} \hskip-1.4mm 
 \stackrel{15}{|}\hskip-1.4mm {0} \hskip-1.4mm 
 \stackrel{16}{|}\hskip-1.4mm {0} \hskip-1.4mm 
 \stackrel{17}{|}\hskip-1.4mm {0} \hskip-1.4mm 
 \stackrel{18}{|}\hskip-1.4mm {0} \hskip-1.4mm 
 \stackrel{19}{|}\hskip-1.4mm {0} \hskip-1.4mm 
 \stackrel{20}{|}, \\
 \hat{E}^2(\sharp) &= & \stackrel{0}{|}\hskip-0.8mm {0} \hskip-0.8mm 
 \stackrel{1}{|}\hskip-0.8mm {0} \hskip-0.8mm 
 \stackrel{2}{|}\hskip-0.8mm {1} \hskip-0.8mm 
 \stackrel{3}{|}\hskip-0.8mm {1} \hskip-0.8mm 
 \stackrel{4}{|}\hskip-0.8mm {1} \hskip-0.8mm 
 \stackrel{5}{|}\hskip-0.8mm {0} \hskip-0.8mm 
 \stackrel{6}{|}\hskip-0.8mm {0} \hskip-0.8mm 
 \stackrel{7}{|}\hskip-0.8mm {1} \hskip-0.8mm 
 \stackrel{8}{|}\hskip-0.8mm {1} \hskip-0.8mm 
 \stackrel{9}{|}\hskip-0.8mm {0} \hskip-1.4mm 
 \stackrel{10}{|}\hskip-1.4mm {0} \hskip-1.4mm 
 \stackrel{11}{|}\hskip-1.4mm {0} \hskip-1.4mm 
 \stackrel{12}{|}\hskip-1.4mm {0} \hskip-1.4mm 
 \stackrel{13}{|}\hskip-1.4mm {0} \hskip-1.4mm 
 \stackrel{14}{|}, \\
 \hat{E}^3(\sharp) &= & \stackrel{0}{|}\hskip-0.8mm {0} \hskip-0.8mm 
 \stackrel{1}{|}\hskip-0.8mm {0} \hskip-0.8mm 
 \stackrel{2}{|}\hskip-0.8mm {1} \hskip-0.8mm 
 \stackrel{3}{|}\hskip-0.8mm {1} \hskip-0.8mm 
 \stackrel{4}{|}\hskip-0.8mm {0} \hskip-0.8mm 
 \stackrel{5}{|}\hskip-0.8mm {1} \hskip-0.8mm 
 \stackrel{6}{|}\hskip-0.8mm {0} \hskip-0.8mm 
 \stackrel{7}{|}\hskip-0.8mm {0} \hskip-0.8mm 
 \stackrel{8}{|}\hskip-0.8mm {0} \hskip-0.8mm 
 \stackrel{9}{|}\hskip-0.8mm {0} \hskip-1.4mm 
 \stackrel{10}{|}, \\
 \hat{E}^4(\sharp) &= & \stackrel{0}{|}\hskip-0.8mm {0} \hskip-0.8mm 
 \stackrel{1}{|}\hskip-0.8mm {0} \hskip-0.8mm 
 \stackrel{2}{|}\hskip-0.8mm {1} \hskip-0.8mm 
 \stackrel{3}{|}\hskip-0.8mm {0} \hskip-0.8mm 
 \stackrel{4}{|}\hskip-0.8mm {0} \hskip-0.8mm 
 \stackrel{5}{|}\hskip-0.8mm {0} \hskip-0.8mm 
 \stackrel{6}{|}, \\
 \hat{E}^5(\sharp) &= & \stackrel{0}{|}\hskip-0.8mm {0} \hskip-0.8mm 
 \stackrel{1}{|}\hskip-0.8mm {0} \hskip-0.8mm 
 \stackrel{2}{|}\hskip-0.8mm {0} \hskip-0.8mm 
 \stackrel{3}{|}\hskip-0.8mm {0} \hskip-0.8mm 
 \stackrel{4}{|}.
\end{eqnarray*}
In $\hat{E}(\sharp)$ the positions of $0$-soliton are $4,\ 7$ and $15$, 
in $\hat{E}^2(\sharp)$ the position of $0$-soliton is $10$, 
in $\hat{E}^3(\sharp)$ there are no $0$-solitons, and so on. 
Note that there is indeterminacy in constructing the original state from the reduced sequence;
for example, both $0110001$ and $1100010$ 
turn to the same reduced state with the same positions of $0$-solitons by $10$-elimination.
Hence we cannot necessarily determine the exact position of the $10$ pair in the original state 
from the position of corresponding $0$-solitons in the reduced state. 


In the $k$-reduced states, $\hat{E}^k({\cal S})$, 
the number of positions of $0$-solitons is $p_{k}-p_{k+1}\enskip (p_{m+1}:=0)$.
Hence there appear $\DIS \sum_{j=1}^s n_j=p_1$ $0$-solitons in total in the reduced states.
We also find that $0$-solitons appear only at the $L_j$-reduced states 
$\hat{E}^{L_j}({\cal S})\ (j=1,2,\ldots ,s)$ and that the number of $0$-solitons is 
$n_j$ in $\hat{E}^{L_j}({\cal S})$. 
There is no $0$-soliton in the other reduced states.
So we denote by $x_{j}^{(k)}\ (k=1,2,\ldots,n_j)$ 
the position of the $k$th $0$-soliton in the $L_j$-reduced state.
Since $\hat{E}^{L_1} ({\cal S})$ only consists of $N-2M$ $0$s, 
we can reconstruct the original state up to some shift from $\{x_k^{(j)}\}_{j=1,}^{s,}{}_{k=1}^{n_j}$.
If, however, we know the original position in the $0$-reduced state of one of the $0$-solitons 
in the $L_1$-reduced state, we can recover the original state, 
because we have only to shift the state obtained by successive insertion of $10$ pairs 
so that it coincides the original position.

For this purpose, it is more convenient to introduce a set of variables 
$\{\alpha_k^{(j)} \}_{j=1,}^{s,}{}_{k=1}^{n_j}$ and $X_{s+1}$. 
First we define the reference positions $X_j\ (j=1,2,\ldots,s+1)$ 
which is defined from $x_1^{(1)}$ recursively as follows.
\begin{itemize}
\item $\tilde{X}_{L_1}=x_1^{(1)}$.
\item We denote by ${\tilde{X}}_{L_1-1}$ the position inbetween $1$ and $0$ 
      in the $(L_1-1)$-reduced state of the $10$ pair which turns into the $0$-soliton 
      at ${\tilde{X}}_{L_1}$ in the $L_1$-reduced state.
\item Similarly we denote by ${\tilde{X}}_{L_1-2}$ the position inbetween $1$ and $0$
      in the $(L_1-2)$-reduced state of the $10$ pair which turns into the position  
      ${\tilde{X}}_{L_1-1}$ in the $(L_1-1)$-reduced state. 
      If there are more than one $10$ pairs inserted at the position, ${\tilde{X}}_{L_1-2}$ 
      is the left most position among them.
\item Repeat the above procedure and obtain $\DIS \tilde{X}_k\ (k=0,1,\ldots ,L_1)$.
\item $\DIS X_j:={\tilde{X}}_{L_j}\quad (j=1,2,\ldots,s+1,\ \mbox{where}\ L_{s+1}:=0)$.
\end{itemize}
In the above example $(\sharp)$, $\tilde{X}_{L_1}=\tilde{X}_5=2,\ \tilde{X}_4=3,\ \tilde{X}_3=4,\ \tilde{X}_2=5,\ 
\tilde{X}_1=6$ and $\tilde{X}_0=9$. Hence $X_1=2,\ X_2=3,\ X_3=5,\ X_4=6$ and $X_5=9$. 
In the terminology of Ref.~\cite{YYT}, $X_j$ is the position of one of the `largest soliton' 
in the $L_j$-reduce state which turns to a $0$-soliton in the $L_1$-reduced state.
Hence $X_{s+1}$ is `the original position in the 0-reduced state' of the 0-soliton at the position $x_1^{(1)}$.

Then we define $\alpha_k^{(j)}\ (1 \le \alpha_k^{(j)} \le N_j)$ by 
\begin{equation}
 \alpha_k^{(j)}= X_j-x_k^{(j)}  \quad \mbox{mod $N_j$} \quad (j=1,2,\ldots, s,\enskip k=1,2,\ldots,n_j),
\end{equation}
where 
\[
N_j:=N-2M+\sum_{i=1}^j2n_i(L_i-L_j) \qquad (j=1,2,\ldots,s)
\]
is the number of entries in the $L_j$-reduced state.
Note that $\alpha_1^{(1)}=N_1$ and $\alpha_k^{(j)}$ is the distance of the $k$th $0$-soliton 
from the position of the `largest soliton' in the $L_j$-reduce state, 
which turns to a $0$-soliton in the $L_1$-reduced state.
Since the state ${\cal S}$ can be determined up to shift by inserting $10$ pairs 
at the positions of $0$-solitons and that between consecutive $1$ and $0$, 
and $X_{s+1}$ determines the amount of the shift, the state ${\cal S}$ is uniquely determined 
by the variables $\{\alpha_k^{(j)}\}_{j=1,}^{s,}{}_{k=1+\delta_{1,j}}^{n_j}$ and $X_{s+1}$. 
Formally we may write that 
\[
\{\alpha_k^{(j)}\}_{k=1}^{n_j} \in S^{n_j}(\Z_{N_j}) 
:=\underbrace{\Z_{N_j}\times \Z_{N_j} \times \cdots \times \Z_{N_j}}_{n_j}/S^{n_j} \quad (j=2,3,\ldots,s),
\]
where $\Z_{N_j}$ is the cyclic group of order $N_j$, $S^{n_j}$ is the symmetric group of order $n_j$, 
and, since there are $N_1+1$ distinct positions\footnote[1]{When another $0$-soliton is located 
at position $X_1$, it is either at the left of the reference $0$-soliton or to the right of it.}
for $(n_1-1)$ 0-solitons in the $L_1$-reduced state,
\[
\{\alpha_k^{(1)}\}_{k=2}^{n_1} \in S^{n_1-1}(\Z_{N_1+1}),
\]
and $X_{s+1}\in \Z_N$.
If we define
\[
\tilde{V}_Y:=S^{n_1-1}(\Z_{N_1+1})\times S^{n_2}(\Z_{N_2})\times \cdots \times S^{n_s}(\Z_{N_s})\times \Z_N,
\]
then an element of $\tilde{V}_Y$ naturally corresponds to a state of the PBBS.
Since there are $n_1$ choices of the reference position, there are exactly $n_1$ elements in $\tilde{V}_Y$ which correspond to a state of the PBBS.
We regard that one element in $\tilde{V}_Y$ is equivalent with another if and only if they corresponds to the same state of the PBBS.
Denoting by $V_Y$ the quotient set of $\tilde{V}_Y$ according to this equivalence relation, 
we obtain the following theorem;
\begin{Theorem}
\label{Bijection} 
Denote by $\Omega_Y$ a set of the states of the PBBS of $M$ balls and $N$ boxes 
with conserved quantities $\{ L_j,\ n_j\}_{j=1}^s$ characterized by the Young diagram $Y$.
Then there is a one to one correspondence between an element of $\Omega_Y$ and that of $V_Y$, 
hence $\DIS \Omega_Y \cong V_Y$.
The explicit bijection is given by the 10-eliminations and its inverse operations with shift
using the variables $X_{s+1}$ and $\{\alpha_k^{(j)}\}_{j=1,}^{s,}{}_{k=1+\delta_{1,j}}^{n_j}$.
\end{Theorem}
As for the explicit construction of a state of the PBBS from an element in $V_Y$, 
see the example given below.

Now we consider the initial value problem of the PBBS.
Let ${\cal S}(t)$ be the state evolved from the state ${\cal S}$ by $t$ time steps. 
From Theorem~\ref{Bijection}, we have that the dynamics of the PBBS can be described by 
an element in $V_Y$.
Hence, to determine a state ${\cal S}(t)$, it is enough to obtain the variables 
$\left( \{\alpha_k^{(j)}(t) \}_{j=1,}^{s,}{}_{k=1+\delta_{1,j}}^{n_j},\ X_{s+1}(t) \right) \in V_Y$ 
from the initial values $\DIS \left( \{\alpha_k^{(j)}(0) \}_{j=1,}^{s,}{}_{k=1+\delta_{1,j}}^{n_j},\ X_{s+1}(0) \right) 
\in V_Y$.
However, the time dependence of these variables has already been given in Ref.~\cite{YYT}: 

\begin{Proposition}[\cite{YYT} Theorem~3.1, Lemma~4.2 and Lemma~4.3]
\label{movement}
For $i=1,2,\ldots ,s+1$, let $\gamma_i(t)$ be $\gamma_1(t):=0,\ \gamma_2(t) :=(L_1-L_2)t$ and 
\begin{equation*}
 \gamma_i(t) :=(L_1-L_i)t +2\sum_{j=2}^{i-1} (L_j-L_i)\sum_{k=1}^{n_j} \beta_{k}^{(j)}(t)
\end{equation*}
for $i=3,4,\ldots ,s+1$, where
\begin{equation*}
 \beta_{k}^{(j)}(t) :=\left\lfloor \frac{\gamma_{j}(t)+\alpha_{k}^{(j)}(0)-1}{N_{j}} \right\rfloor 
 \qquad (j=2,3,\ldots ,s,\enskip k=1,2,\ldots ,n_j).
\end{equation*}
Then it holds that 
\begin{equation*}
 \alpha_{k}^{(j)}(t) =\alpha_{k}^{(j)}(0) +\gamma_j(t)\quad \mod N_j.
\end{equation*}
and
\begin{equation*}
 X_{s+1}(t) =X_{s+1}(0) +\gamma_{s+1}(t)\quad \mod N. 
\end{equation*}
\end{Proposition}

\begin{Remark}
Note that both $\gamma_i(t)$s and $\alpha_{k}^{(j)}(t)$s are determined recursively.
As we shall see, practically we have only to use the relation
\begin{equation*}
 \alpha_{k}^{(j)}(0) +\gamma_j(t) = N_j\beta_{k}^{(j)}(t) +\alpha_{k}^{(j)}(t).
\end{equation*}
\end{Remark}

In conclusion, we have solved the initial value problem of the PBBS which may be stated as 
\begin{Theorem}
\label{Initial}
The initial value problem of the PBBS is solved in the space of $V_Y$.
Its dynamics is explicitly given in Proposition~\ref{movement}.
\end{Theorem}


In the rest of this article, we explain how to obtain the time evolution of a state of the PBBS
by means of an example.
Suppose that we have the state $(\sharp)$ which we have used previously
\begin{equation*}
 (\sharp) \qquad 00111011100100011110001101000000
\end{equation*}
at time step $t=0$.
From Fig.~\ref{YoungD} and $\hat{E}^k(\sharp)$, we find $s=4,\ N_1=4,\ N_2=6,\ N_3=14,\ N_4=20$.
Then referring to the reduced sequences $\hat{E}^k(\sharp)$, we obtain
\begin{equation*}
 x_1^{(1)}(0)=2;\enskip x_1^{(2)}(0)=3;\enskip x_1^{(3)}(0)=10;\enskip 
 x_1^{(4)}(0)=4,\ x_2^{(4)}(0)=7,\ x_3^{(4)}(0)=15.
\end{equation*}
The reference positions $X_j(0)$ $(j=1,2,\ldots,5)$ are found to be 
\[
 X_1(0)=2,\ X_2(0)=3,\ X_3(0)=5,\ X_4(0)=6,\ X_5(0)=9.
\]
Hence $\alpha_k^{(j)}(0)$s are given as
\begin{equation*}
 \alpha_1^{(1)}(0)=4;\enskip \alpha_1^{(2)}(0)=6;\enskip \alpha_1^{(3)}(0)=9;\enskip 
 \alpha_1^{(4)}(0)=2,\ \alpha_2^{(4)}(0)=19,\ \alpha_3^{(4)}(0)=11.
\end{equation*}
Now let us consider the state at $t=10000$. 
According to Proposition~\ref{movement}, $\alpha_k^{(j)}(t)$s and $\gamma_i(t)$s are
calculated recursively as follows;
\begin{eqnarray*}
 \mbox{(1)} && N_2 =6,\ 
               \gamma_2(t) =(L_1-L_2)t =10000 \\
  && \quad \Rightarrow \enskip 
      \alpha_1^{(2)}(0) +\gamma_2(t) =6+10000 =1667\cdot 6 +4 \\
  && \qquad \Rightarrow \enskip 
        \alpha_1^{(2)}(t) =4; \\
 \mbox{(2)} && N_3 =14,\ 
               \gamma_3(t) =(L_1-L_3)\cdot t +2(L_2-L_3)\cdot 1667 =36668 \\
  && \quad \Rightarrow \enskip 
      \alpha_1^{(3)}(0) +\gamma_3(t) =9+36668 =2619\cdot 14 +11 \\
  && \qquad \Rightarrow \enskip 
        \alpha_1^{(3)}(t) =11; \\
 \mbox{(3)} && N_4 =20,\\
            && \gamma_4(t) =(L_1-L_4)\cdot t +2(L_2-L_4)\cdot 1667 +2(L_3-L_4)\cdot 2619 =55240 \\
  && \quad \Rightarrow \enskip \left\{ \begin{array}{l}
      \alpha_1^{(4)}(0) +\gamma_4(t) =2+55240  =2762\cdot 20 +2  \\
      \alpha_2^{(4)}(0) +\gamma_4(t) =19+55240 =2762\cdot 20 +19 \\
      \alpha_3^{(4)}(0) +\gamma_4(t) =11+55240 =2762\cdot 20 +11 \\
     \end{array} \right. \\
  && \qquad \Rightarrow \enskip \left\{ \begin{array}{l}
        \alpha_1^{(4)}(t) =2,  \\
        \alpha_2^{(4)}(t) =19, \\
        \alpha_3^{(4)}(t) =11; \\
     \end{array} \right. \\
 \mbox{(4)} && N =32,\\ 
  && \gamma_5(t) =(L_1-L_5)\cdot t +2(L_2-L_5)\cdot 1667 +2(L_3-L_5)\cdot 2619 \\
  && \qquad \qquad 
                  +2(L_4-L_5)\cdot (2762+2762+2762) =90384 \\
  && \quad \Rightarrow \enskip X_5(t) =X_5(0) +\gamma_5(t) \mod N =9 +90384 \mod 32 \\
  && \qquad \qquad \quad \enskip =25. 
\end{eqnarray*}
From these data, the state at $t=10000$ is constructed up to shift by inserting $10$ pairs as;
\begin{eqnarray*}
 && \stackrel{0}{|}\hskip-0.8mm {0} \hskip-0.8mm 
 \stackrel{1}{|}\hskip-0.8mm {0} \hskip-0.8mm 
 \stackrel{2}{|}\hskip-0.8mm {0} \hskip-1.2mm 
 \stackrel{3^{o}}{|}\hskip-1.2mm {0} \hskip-0.8mm 
 \stackrel{4}{|}, \\
 &\longrightarrow & \stackrel{0^{o}}{|}\hskip-1.2mm {0} \hskip-0.8mm 
 \stackrel{1}{|}\hskip-0.8mm {0} \hskip-0.8mm 
 \stackrel{2}{|}\hskip-0.8mm {0} \hskip-0.8mm 
 \stackrel{3}{|}\hskip-0.8mm {1} \hskip-1.2mm 
 \stackrel{4^*}{|}\hskip-1.2mm {0} \hskip-0.8mm 
 \stackrel{5}{|}\hskip-0.8mm {0} \hskip-0.8mm 
 \stackrel{6}{|}, \\
 &\longrightarrow & \stackrel{0}{|}\hskip-0.8mm {1} \hskip-0.8mm 
 \stackrel{1}{|}\hskip-0.8mm {0} \hskip-0.8mm 
 \stackrel{2}{|}\hskip-0.8mm {0} \hskip-0.8mm 
 \stackrel{3}{|}\hskip-0.8mm {0} \hskip-0.8mm 
 \stackrel{4}{|}\hskip-0.8mm {0} \hskip-0.8mm 
 \stackrel{5}{|}\hskip-0.8mm {1} \hskip-0.8mm 
 \stackrel{6}{|}\hskip-0.8mm {1} \hskip-1.2mm 
 \stackrel{7^*}{|}\hskip-1.2mm {0} \hskip-0.8mm 
 \stackrel{8}{|}\hskip-0.8mm {0} \hskip-0.8mm 
 \stackrel{9}{|}\hskip-0.8mm {0} \hskip-1.4mm 
 \stackrel{10}{|}, \\
 &\longrightarrow & \stackrel{0}{|}\hskip-0.8mm {1} \hskip-0.8mm 
 \stackrel{1}{|}\hskip-0.8mm {1} \hskip-0.8mm 
 \stackrel{2}{|}\hskip-0.8mm {0} \hskip-0.8mm 
 \stackrel{3}{|}\hskip-0.8mm {0} \hskip-0.8mm 
 \stackrel{4}{|}\hskip-0.8mm {0} \hskip-0.8mm 
 \stackrel{5}{|}\hskip-0.8mm {0} \hskip-0.8mm 
 \stackrel{6}{|}\hskip-0.8mm {0} \hskip-0.8mm 
 \stackrel{7}{|}\hskip-0.8mm {1} \hskip-0.8mm 
 \stackrel{8}{|}\hskip-0.8mm {1} \hskip-0.8mm 
 \stackrel{9}{|}\hskip-0.8mm {1} \hskip-1.8mm 
 \stackrel{10^*}{|}\hskip-1.8mm {0} \hskip-1.4mm 
 \stackrel{11}{|}\hskip-1.4mm {0} \hskip-1.4mm 
 \stackrel{12}{|}\hskip-1.4mm {0} \hskip-1.4mm 
 \stackrel{13^{o}}{|}\hskip-1.4mm {0} \hskip-1.4mm 
 \stackrel{14}{|}, \\
 &\longrightarrow & \stackrel{0}{|}\hskip-0.8mm {1} \hskip-0.8mm 
 \stackrel{1}{|}\hskip-0.8mm {1} \hskip-1.2mm 
 \stackrel{2^{o}}{|}\hskip-1.2mm {1} \hskip-0.8mm 
 \stackrel{3}{|}\hskip-0.8mm {0} \hskip-0.8mm 
 \stackrel{4}{|}\hskip-0.8mm {0} \hskip-0.8mm 
 \stackrel{5}{|}\hskip-0.8mm {0} \hskip-0.8mm 
 \stackrel{6}{|}\hskip-0.8mm {0} \hskip-0.8mm 
 \stackrel{7}{|}\hskip-0.8mm {0} \hskip-0.8mm 
 \stackrel{8}{|}\hskip-0.8mm {0} \hskip-0.8mm 
 \stackrel{9}{|}\hskip-0.8mm {1} \hskip-1.4mm 
 \stackrel{10}{|}\hskip-1.4mm {1} \hskip-1.8mm 
 \stackrel{11^{o}}{|}\hskip-1.8mm {1} \hskip-1.4mm 
 \stackrel{12}{|}\hskip-1.4mm {1} \hskip-1.8mm 
 \stackrel{13^*}{|}\hskip-1.8mm {0} \hskip-1.8mm 
 \stackrel{14^{o}}{|}\hskip-1.8mm {0} \hskip-1.4mm 
 \stackrel{15}{|}\hskip-1.4mm {0} \hskip-1.4mm 
 \stackrel{16}{|}\hskip-1.4mm {0} \hskip-1.4mm 
 \stackrel{17}{|}\hskip-1.4mm {1} \hskip-1.4mm 
 \stackrel{18}{|}\hskip-1.4mm {0} \hskip-1.4mm 
 \stackrel{19}{|}\hskip-1.4mm {0} \hskip-1.4mm 
 \stackrel{20}{|}, \\
 &\longrightarrow & \stackrel{0}{|}\hskip-0.8mm {1} \hskip-0.8mm 
 \stackrel{1}{|}\hskip-0.8mm {1} \hskip-0.8mm 
 \stackrel{2}{|}\hskip-0.8mm {1} \hskip-0.8mm 
 \stackrel{3}{|}\hskip-0.8mm {0} \hskip-0.8mm 
 \stackrel{4}{|}\hskip-0.8mm {1} \hskip-0.8mm 
 \stackrel{5}{|}\hskip-0.8mm {1} \hskip-0.8mm 
 \stackrel{6}{|}\hskip-0.8mm {0} \hskip-0.8mm 
 \stackrel{7}{|}\hskip-0.8mm {0} \hskip-0.8mm 
 \stackrel{8}{|}\hskip-0.8mm {0} \hskip-0.8mm 
 \stackrel{9}{|}\hskip-0.8mm {0} \hskip-1.4mm 
 \stackrel{10}{|}\hskip-1.4mm {0} \hskip-1.4mm 
 \stackrel{11}{|}\hskip-1.4mm {0} \hskip-1.4mm 
 \stackrel{12}{|}\hskip-1.4mm {0} \hskip-1.4mm 
 \stackrel{13}{|}\hskip-1.4mm {1} \hskip-1.4mm 
 \stackrel{14}{|}\hskip-1.4mm {1} \hskip-1.4mm 
 \stackrel{15}{|}\hskip-1.4mm {1} \hskip-1.4mm 
 \stackrel{16}{|}\hskip-1.4mm {0} \hskip-1.4mm 
 \stackrel{17}{|}\hskip-1.4mm {1} \hskip-1.4mm 
 \stackrel{18}{|}\hskip-1.4mm {1} \hskip-1.4mm 
 \stackrel{19}{|}\hskip-1.4mm {1} \hskip-1.8mm 
 \stackrel{20^*}{|}\hskip-1.8mm {0} \hskip-1.4mm 
 \stackrel{21}{|}\hskip-1.4mm {0} \hskip-1.4mm 
 \stackrel{22}{|}\hskip-1.4mm {1} \hskip-1.4mm 
 \stackrel{23}{|}\hskip-1.4mm {0} \hskip-1.4mm 
 \stackrel{24}{|}\hskip-1.4mm {0} \hskip-1.4mm 
 \stackrel{25}{|}\hskip-1.4mm {0} \hskip-1.4mm 
 \stackrel{26}{|}\hskip-1.4mm {0} \hskip-1.4mm 
 \stackrel{27}{|}\hskip-1.4mm {1} \hskip-1.4mm 
 \stackrel{28}{|}\hskip-1.4mm {1} \hskip-1.4mm 
 \stackrel{29}{|}\hskip-1.4mm {0} \hskip-1.4mm 
 \stackrel{30}{|}\hskip-1.4mm {0} \hskip-1.4mm 
 \stackrel{31}{|}\hskip-1.4mm {0} \hskip-1.4mm 
 \stackrel{32}{|}.
\end{eqnarray*}
Here  `$\DIS \ \stackrel{j^*}{|}\ $' denotes the position of the largest soliton (referring position) 
and `$\DIS \ \stackrel{j^o}{|}\ $' denotes that of a 0-soliton.
The position $x_1^{(1)}(t)$ can be chosen arbitrary, and we took $x_1^{(1)}(t)=3$ in this example.
Finally we translate the above state so that the position of the largest soliton coincides 
with $X_{s+1}(t)$ as
\begin{equation*}
 \stackrel{0}{|}\hskip-0.8mm {1} \hskip-0.8mm 
 \stackrel{1}{|}\hskip-0.8mm {1} \hskip-0.8mm 
 \stackrel{2}{|}\hskip-0.8mm {0} \hskip-0.8mm 
 \stackrel{3}{|}\hskip-0.8mm {0} \hskip-0.8mm 
 \stackrel{4}{|}\hskip-0.8mm {0} \hskip-0.8mm 
 \stackrel{5}{|}\hskip-0.8mm {1} \hskip-0.8mm 
 \stackrel{6}{|}\hskip-0.8mm {1} \hskip-0.8mm 
 \stackrel{7}{|}\hskip-0.8mm {1} \hskip-0.8mm 
 \stackrel{8}{|}\hskip-0.8mm {0} \hskip-0.8mm 
 \stackrel{9}{|}\hskip-0.8mm {1} \hskip-1.4mm 
 \stackrel{10}{|}\hskip-1.4mm {1} \hskip-1.4mm 
 \stackrel{11}{|}\hskip-1.4mm {0} \hskip-1.4mm 
 \stackrel{12}{|}\hskip-1.4mm {0} \hskip-1.4mm 
 \stackrel{13}{|}\hskip-1.4mm {0} \hskip-1.4mm 
 \stackrel{14}{|}\hskip-1.4mm {0} \hskip-1.4mm 
 \stackrel{15}{|}\hskip-1.4mm {0} \hskip-1.4mm 
 \stackrel{16}{|}\hskip-1.4mm {0} \hskip-1.4mm 
 \stackrel{17}{|}\hskip-1.4mm {0} \hskip-1.4mm 
 \stackrel{18}{|}\hskip-1.4mm {1} \hskip-1.4mm 
 \stackrel{19}{|}\hskip-1.4mm {1} \hskip-1.4mm 
 \stackrel{20}{|}\hskip-1.4mm {1} \hskip-1.4mm 
 \stackrel{21}{|}\hskip-1.4mm {0} \hskip-1.4mm 
 \stackrel{22}{|}\hskip-1.4mm {1} \hskip-1.4mm 
 \stackrel{23}{|}\hskip-1.4mm {1} \hskip-1.4mm 
 \stackrel{34}{|}\hskip-1.4mm {1} \hskip-1.8mm 
 \stackrel{25^*}{|}\hskip-1.8mm {0} \hskip-1.4mm 
 \stackrel{26}{|}\hskip-1.4mm {0} \hskip-1.4mm 
 \stackrel{27}{|}\hskip-1.4mm {1} \hskip-1.4mm 
 \stackrel{28}{|}\hskip-1.4mm {0} \hskip-1.4mm 
 \stackrel{29}{|}\hskip-1.4mm {0} \hskip-1.4mm 
 \stackrel{30}{|}\hskip-1.4mm {0} \hskip-1.4mm 
 \stackrel{31}{|}\hskip-1.4mm {0} \hskip-1.4mm 
 \stackrel{32}{|}.
\end{equation*}
This is the state at $t=10000$.

In this article, we have solved the initial value problem in an elementary way.
We also remark that our method can be equally applied to extended PBBSs with carrier 
capacity $\ell$ as those treated in Ref.~\cite{MIT4}. 
In these PBBSs, we have only to replace $L_j \to \min[L_j,\ell]$ in Proposition \ref{movement}
and apply the above procedures. 
Clarifying the relation between our methods and previous work based 
on algebraic curves and representation theories, is one of the important problems 
we want to address in the future.

\hbreak
\hbreak
\noindent
{\large  \bf Acknowledgments}

The authors are grateful to Prof. R. Willox for helpful comments on the present work. 
This work is supported in part by the 21st century COE program 
at the Graduate School of Mathematical Sciences of the University of Tokyo. 



\end{document}